\input{aipcheck}

\documentclass[
    ,final            
  ]
  {aipproc}

\layoutstyle{6x9}

\begin{document}

\title{Survey for Supernovae in Massive High-Redshift Clusters}

\classification{}
\keywords      {Supernovae}

\author{Keren Sharon}{
  address={School of Physics and Astronomy, Tel Aviv University}
}
\author{Avishay Gal-Yam}{
  address={Division of Physics, Mathematics and Astronomy, California Institute of Technology}
  ,altaddress={Hubble Fellow} 
}

\author{Dan Maoz}{
  address={School of Physics and Astronomy, Tel Aviv University}
}

\author{Megan Donahue}{
  address={Michigan State University}
}
\author{Harald Ebeling}{
  address={University of Hawaii}
}
\author{Richard S. Ellis}{
  address={Division of Physics, Mathematics and Astronomy, California Institute of Technology}
}

\author{Alex V. Filippenko}{
  address={University of California - Berkeley}
}
\author{Ryan Foley}{
  address={University of California - Berkeley}
}
\author{Wendy L. Freedman}{
  address={Carnegie Institution of Washington}
}
\author{Robert P. Kirshner}{
  address={Harvard University}
}
\author{Jean-Paul Kneib}{
  address={Observatoire de Marseille}
}
\author{Thomas Matheson}{
  address={National Optical Astronomy Observatories, AURA}
}
\author{John S. Mulchaey}{
  address={Carnegie Institution of Washington}
}
\author{Vicki L. Sarajedini}{
  address={University of Florida}
}
\author{Mark Voit}{
  address={Michigan State University}
}

\begin{abstract}
We describe our ongoing program designed to measure the SN-Ia rate in a sample of massive z=0.5-0.9 galaxy clusters. The SN-Ia rate is a poorly known observable, especially at high z, and in cluster environments. The SN rate and its redshift dependence can serve as powerful discrimiminants for a number of key issues in astrophysics and cosmology. Our observations will put clear constraints on the characteristic SN-Ia ``delay time'', the typical time between the formation of a stellar population and the explosion of some of its members as SNe-Ia. Such constraints can exclude entire categories of SN-Ia progenitor models, since different models predict different delays. These data will
also help to resolve the question of the dominant source of the high metallicity in the intracluster medium (ICM) - SNe-Ia, or core-collapse SNe from an early stellar population with a top-heavy IMF, perhaps those population III stars responsible for the early re-ionization of the Universe. Since clusters are excellent laboratories for studying enrichment (they generally have a simple star-formation history, and matter cannot leave their deep potentials), the results will be relevant for understanding metal enrichment in general, and the possible role of first generation stars in early Universal enrichment. Observations obtained so far during cycles 14 and 15 yield many SNe in our cluster fields, but our follow-up campaign reveals most are not in cluster galaxies.

\end{abstract}

\maketitle

\section{Preliminary Results}
We were allocated 30 orbits of $HST$ time during cycles 14 and 15\footnote{see \url{http://www.astro.caltech.edu/~avishay/hst.html} for additional details}, 
to re-image 15 X-ray luminous clusters, which were already imaged by ACS in the past (e.g., Fig. 1). 
We compare our new images with previous data, and promptly search for transients using image subtraction (Fig. 2). 

Follow-up spectroscopy (e.g., Fig. 3) is used to determine host redshifts for faint SNe (typical events have $I>25$ mag) and thus reject forground or background events and AGN. Color imaging is used to further classify SNe based on their colors and luminosities. Partial follow-up has already been secured, and additional Keck follow-up has 
been allocated. Table~1 lists the SN candidates that were found in Cycle 14. For each event we give an identification (if possible). Two cluster SNe Ia have been secured so far, and additional 2-6 (hostless, or in galaxies which appear to be cluster members based on color and {\it HST} morphology) should also contribute to the cluster rate. All other events appear to reside in galaxies which are most likely interlopers and AGN.

\begin{figure}[b]
\includegraphics{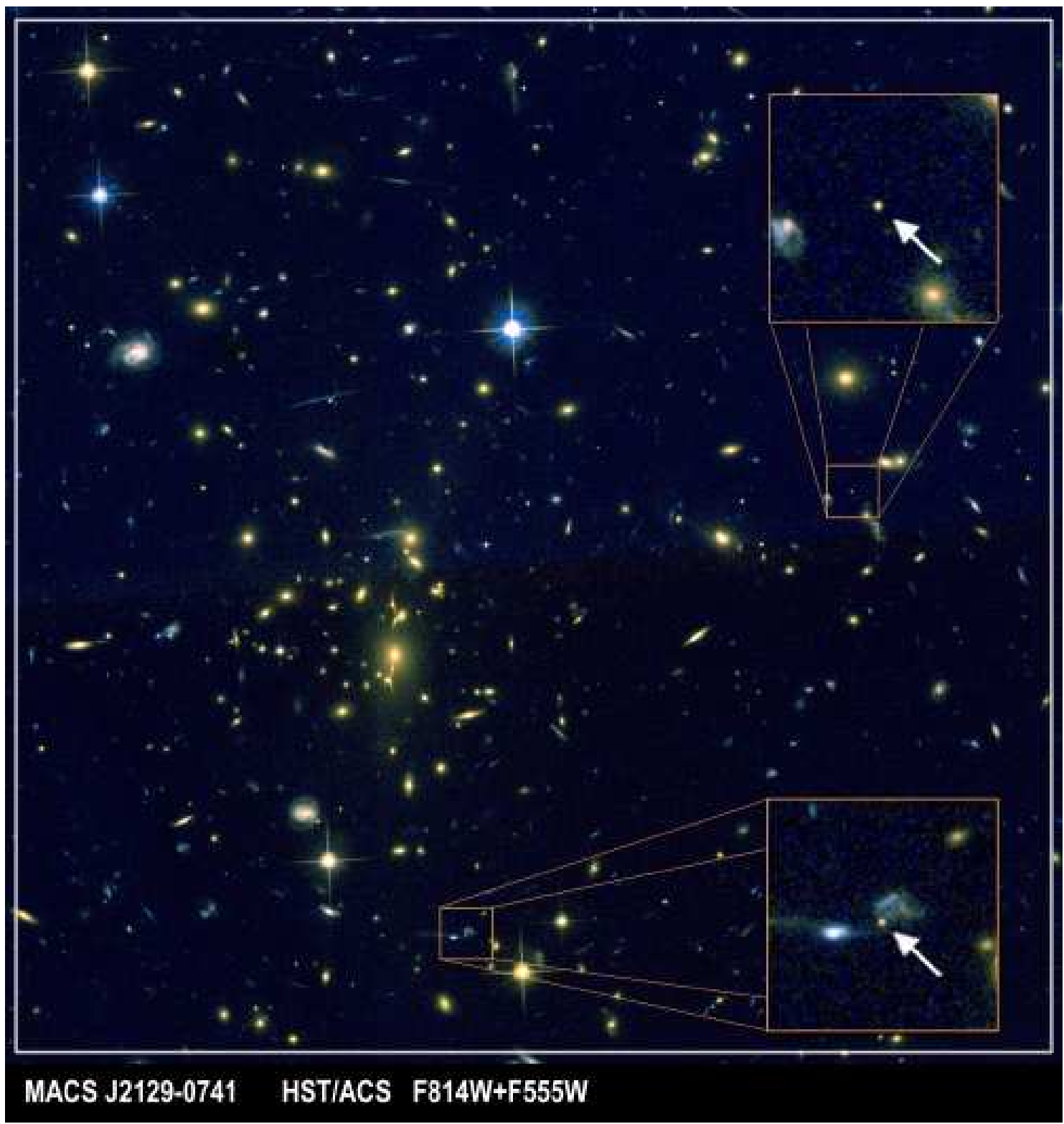}
\caption{
Example of an ACS cluster image from our ongoing program,
with zoom-ins on two of the SNe we have found.
}
\label{snrates} 
\end{figure}

\begin{figure}
  \includegraphics{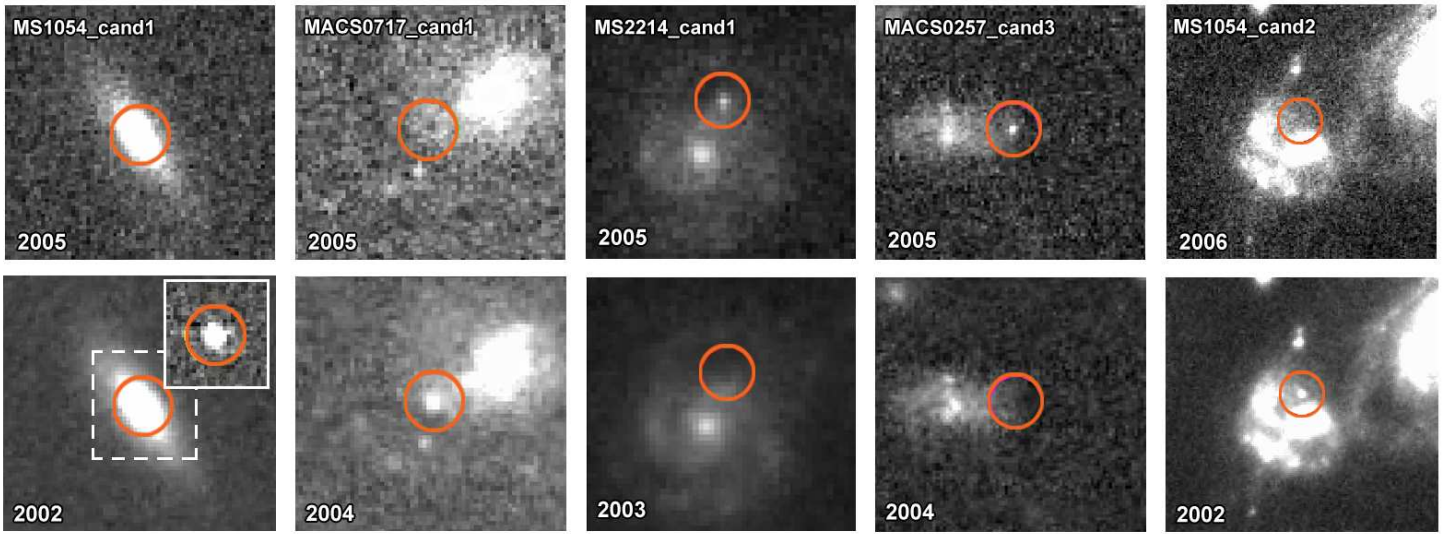}
  \caption{Examples of SNe and SN candidates found in our ongoing program. Keck spectra of the host of candidate-1 in MS1054 ($z=0.83$) show it is an early-type cluster galaxy devoid of any emission lines, arguing that the transient from 2002 (its appearance in the difference image is shown as an inset) was a near-nuclear SN-Ia in a cluster-galaxy. The other events shown have similarly been classified as cluster, background and foreground events (Table 1).}
\end{figure}

\begin{figure}
  \includegraphics{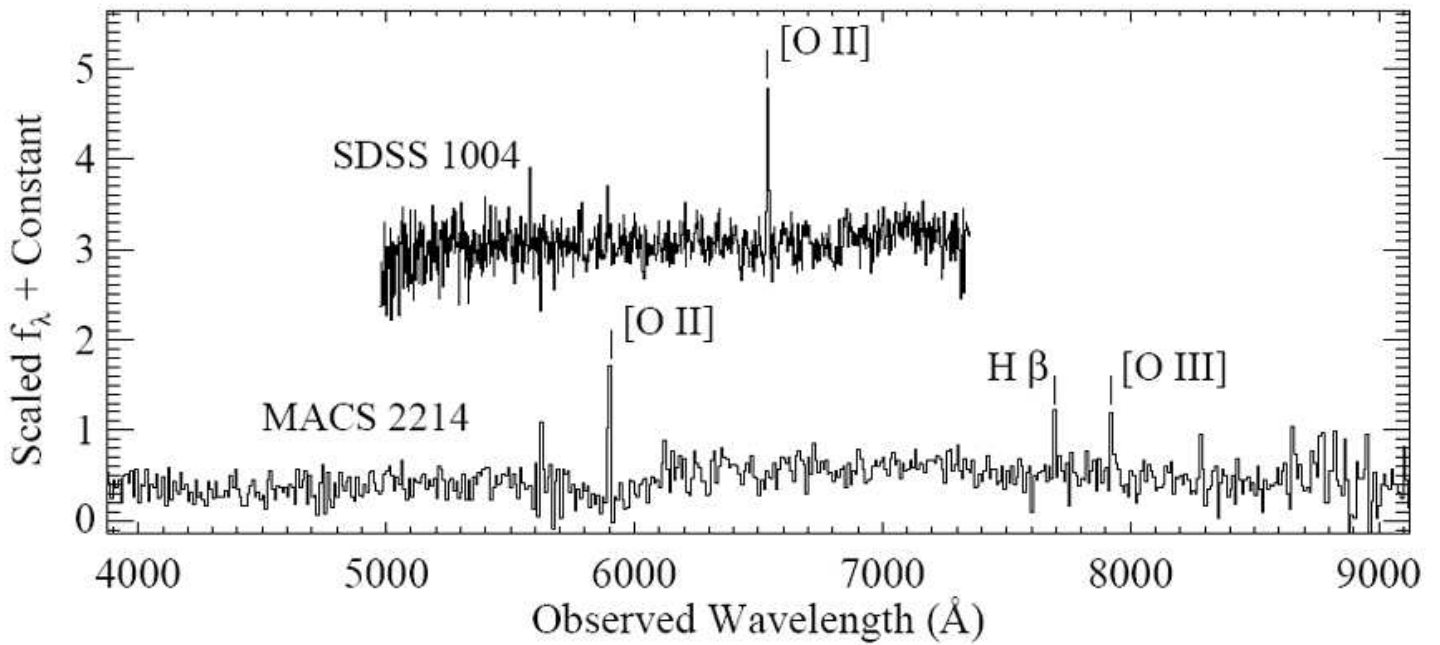}
  \caption{Examples of recently obtained Keck spectra. Top: a faint SN candidate, the spectrum shown contains the sum of the SN and host light, and final extraction awaits galaxy host light subtraction after the SN has faded. Bottom: a SN host galaxy, the derived redshift, $z=0.582$, shows this was a background event.}
\end{figure}

\begin{table}
\begin{tabular}{llll}
\hline
\tablehead{1}{l}{b}{SN} &
\tablehead{1}{l}{b}{Magnitude} &
\tablehead{1}{l}{b}{Redshift} &
\tablehead{1}{l}{b}{ID}  \\
\hline
M0257\_1 & 25.3 & ? &\\
M0257\_2 & 27.4 & ? & Hostless SN\\
M0257\_3 & 26.1 & 0.73 &Background SN\\
M0257\_4 & 26.4 & ? &\\
M0257\_5 & 25.7 & ? &\\
M0647\_1 & 25.8 & ? & \\
M0647\_2 & 26.8 & ? & \\
M0717\_1 & 25.8 & 0.55 & Cluster, Non-Ia SN\\
M0911\_1 & 24.3 & ? & \\
M0911\_2 & 24.9 & ? & \\
M0911\_3 & 24.0 & ? & \\
M1149\_1 & 23.9 & ? & \\
M1149\_2 & 24.6 & ? & \\
M1149\_3 & 24.7 & ? & \\
M2129\_1 & 24.9 & 0.87 & Background SN\\
M2129\_2 & 24.4 & ? & Hostless SN\\
M2214\_1 & 26.2 & 0.58 & Background SN\\
M2214\_2 &      & ? & \\
M2214\_3 & 24.1 & ? & \\
M2214\_4 &      & ? & \\
C1226\_1 & 25.8 & ? &\\
C1226\_2 & 24.0 & 0.90 & Cluster SN Ia\\
M1054\_1 & 24.8 & ? & \\
M1054\_1 & 23.4 & 0.83 & Cluster SN Ia\\
M0451\_1 & 25.5 & 0.16 & Foreground SN\\
C0152\_1 & 25.2 & 1.27 & AGN\\
C0152\_2 & >24 & ? & \\
\hline
\end{tabular}
\caption{Candidate SNe from our {\it HST} search.}
\label{tab:a}
\end{table}

\begin{theacknowledgments}
A.G. acknowledges support
by NASA through Hubble Fellowship grant \#HST-HF-01158.01-A awarded by STScI, which is operated by AURA, Inc., for NASA, under contract NAS
5-26555. A.G. further acknowledges the hospitality of the community of Cefalu and the efforts of the
organizers of the 2006 Cefalu international astronomy conference, during which this work has come to fruition. 
\end{theacknowledgments}




\end{document}